\documentclass[twocolumn,showpacs,preprintnumbers,amsmath,amssymb]{revtex4}

\usepackage{graphicx}
\usepackage{dcolumn}
\usepackage{bm}

\begin{document}

\preprint{}

\title{Stark-tuned F\"orster resonance and dipole blockade for two to five cold Rydberg atoms: 
Monte Carlo simulations for various spatial configurations}
\author{I.~I.~Ryabtsev}
  \email{ryabtsev@isp.nsc.ru}
\author{D.~B.~Tretyakov}
\author{I.~I.~Beterov}
\author{V.~M.~Entin}
\author{E.~A.~Yakshina}
\affiliation{Institute of Semiconductor Physics\\ Prospekt Lavrentyeva 13, 630090 Novosibirsk, Russia }

\date{November 11, 2010}

\begin{abstract}
Results of numerical Monte Carlo simulations for the Stark-tuned F$\ddot{\mathrm{o}}$rster resonance and dipole blockade 
between two to five cold rubidium Rydberg atoms in various spatial configurations are presented. The effects of the atoms' 
spatial uncertainties on the resonance amplitude and spectra are investigated. The feasibility of observing coherent 
Rabi-like population oscillations at a F$\ddot{\mathrm{o}}$rster resonance between two cold Rydberg atoms is analyzed. Spectra 
and the fidelity of the Rydberg dipole blockade are calculated for various experimental conditions, including nonzero 
detuning from the F$\ddot{\mathrm{o}}$rster resonance and finite laser line width. The results are discussed in the context of 
quantum-information processing with Rydberg atoms.
\end{abstract}

\pacs{32.80.Ee, 03.67.Lx, 34.10.+x, 32.70.Jz}
 \maketitle

\section{INTRODUCTION}

Strong dipole-dipole interaction (DDI) between highly excited Rydberg atoms with a principal quantum number $n\gg1$ is a key element for quantum logic gates with qubits based on single alkali-metal atoms [1,2]. Two basic proposals consider a short-term DDI of two close Rydberg atoms to implement conditional quantum phase gates [3] or laser excitation of only one Rydberg atom in a mesoscopic ensemble (dipole blockade) to obtain entangled qubits [4].

DDI arises as soon as two ground-state atoms are laser excited to the Rydberg states of the opposite parity [2,5]. Alternatively, two Rydberg atoms in an identical state may interact via a F$\ddot{\mathrm{o}}$rster resonance if the atoms are excited to a level that lies midway between two other levels of the opposite parity [6]. Due to DDI one of the atoms undergoes a transition to a higher Rydberg state, while the other atom goes to a lower Rydberg state. This process is resonantly enhanced at zero energy defect. For some Rydberg levels F$\ddot{\mathrm{o}}$rster resonances can be precisely tuned with an electric field via the Stark effect [2,6]. Accidental quasi-F$\ddot{\mathrm{o}}$rster resonances with a small energy defect can also be found for certain Rydberg states [7,8].

Accidental quasi-F$\ddot{\mathrm{o}}$rster resonances are most suitable for the dipole blockade, which is insensitive to decoherence and fluctuations of the DDI strength [1,8]. Dipole blockade directly generates entanglement between qubits. Two recent experiments [9,10] have demonstrated entanglement and controlled-NOT quantum gates for the two optically trapped and individually addressed Rb atoms. Both experiments employed the dipole blockade at the laser excitation to the Rydberg states interacting via quasi-F$\ddot{\mathrm{o}}$rster resonances. The energy defects of a few MHz were small compared to the DDI strength, and blockade efficiency was close to 1. Partial dipole blockade has also been demonstrated in several earlier experiments with quasi-F$\ddot{\mathrm{o}}$rster or even van der Waals interactions between Rydberg atoms [11-14].

An alternative approach to quantum-information processing with Rydberg atoms relies on the conditional quantum phase gates (QPG) [7,15,16]. This approach has the advantage that it does not require individual addressing of the atoms and strong DDI [1], but it requires coherence to control the phase of the collective wave function of the two atoms. Coherent interaction means that Rabi-like population oscillations of high contrast between the initial and final collective states must be obtained. This can be achieved with the resonant DDI, in particular, with Stark-tuned F$\ddot{\mathrm{o}}$rster resonances. Compared to the accidental quasi-F$\ddot{\mathrm{o}}$rster resonances, Stark-tuned resonance is more flexible in controlling the interaction strength. A weak electric field can be used for fast switching of the interaction [17] and laser excitation [18], and for adjusting the phase of the collective wave function by inducing the Rabi-like population oscillations and stopping them at a proper time. Efficient dipole blockade can be implemented with Stark-tuned F$\ddot{\mathrm{o}}$rster resonances, as well. 

QPG or complete dipole blockade at a Stark-tuned F$\ddot{\mathrm{o}}$rster resonance between two or a few Rydberg atoms has not been reported yet. For many atoms, partial dipole blockade was observed [19,20], and experiments on the population dynamics at F$\ddot{\mathrm{o}}$rster resonance have been done [21,22].

In our recent experiment [23] we observed for the first time a Stark-tuned F$\ddot{\mathrm{o}}$rster resonance between two rubidium Rydberg atoms randomly positioned in a small laser excitation volume. High-resolution spectra of the F$\ddot{\mathrm{o}}$rster resonance have been obtained for $N=2-5$ of the detected Rydberg atoms that allowed us to investigate in detail the line shape dependence on \textit{N} and effective coherence time. The related theoretical analysis relied on numerical Monte Carlo simulations for two to five Rydberg atoms. Although the theoretical model was simplified by ignoring the hyperfine structure of the Rydberg levels and the broadenings due to parasitic electromagnetic fields, we have shown that these broadenings can be taken into account by reducing the effective interaction time. This reduction results in a Fourier broadening of the resonance to a value that matches the experimental width. A good agreement between experiment and theory has confirmed the validity of the model used.

In this article we present a more detailed description of our Monte Carlo model and apply it to simulate a possible observation of the coherent Rabi-like population oscillations at a Stark-tuned F$\ddot{\mathrm{o}}$rster resonance. We also simulate numerically the excitation spectra for the dipole blockade effect at a F$\ddot{\mathrm{o}}$rster resonance. The simulations were performed for various spatial configurations of the interacting Rydberg atoms relevant to the typical experimental conditions. The effects of the atoms' spatial uncertainties on the F$\ddot{\mathrm{o}}$rster resonance and dipole blockade are analyzed.

\section{INTERACTION OF TWO RYDBERG ATOMS}

\subsection{Two frozen atoms}

The operator of the dipole-dipole interaction between two atoms \textit{a} and \textit{b} is

\begin{equation} \label{Eq1} 
\hat {V}_{ab} = \frac{{1}}{{4\pi \varepsilon _{0}} }\left[ {\frac{{\hat 
{\mathbf{d}}_{a} \cdot \hat {\mathbf{d}}_{b}} }{{R_{ab}^{3}} } - \frac{{3\,\,\left( {\hat {\mathbf{d}}_{a} 
\cdot \mathbf{R}_{ab}}  \right)\,\left( {\hat {\mathbf{d}}_{b} \cdot \mathbf{R}_{ab}}  \right)}}{{R_{ab}^{5}} }} 
\right].  
\end{equation} 

\noindent Here $\hat {\mathbf{d}}_{a} $ and $\hat {\mathbf{d}}_{b} $ are the dipole-moment operators for the atoms \textit{a} and \textit{b}, $\mathbf{R}_{ab} $ is a vector connecting the two atoms, and $\varepsilon _{0}$ is the dielectric constant. The interaction energy depends on the mutual orientation of the two atom dipoles (defined by their quantum states) and on their orientation with respect to $\mathbf{R}_{ab} $. 

Degenerate atomic states usually produce several Stark-tuned F$\ddot{\mathrm{o}}$rster resonances corresponding to different momentum projections. Nevertheless, various Stark-tuned F$\ddot{\mathrm{o}}$rster resonances in Rydberg atoms should have similar spectroscopic properties. The difference is mainly in the dipole moments that define the interaction strength and in the initial resonance detuning in a zero electric field. Therefore, the calculations can be done for an arbitrary F$\ddot{\mathrm{o}}$rster resonance, whose theoretical description can be scaled later to other dipole moments and initial detunings.

In this article we consider an example of the Stark-tuned F$\ddot{\mathrm{o}}$rster resonance ${\rm Rb}(37P_{3/2} )+{\rm Rb}(37P_{3/2} )\to {\rm Rb}(37S_{1/2} )+{\rm Rb}(38S_{1/2} )$ for two or more Rb Rydberg atoms [see Fig.~1(a) for the level and transition scheme]. The initial energy detuning $[2{\rm E}(37P_{3/2} )-{\rm E}(37S_{1/2} )-{\rm E}(38S_{1/2} )]/h$ in a zero electric field is 103 MHz. The proper choice of the exciting laser polarization (along the dc electric field) provides excitation of only 37\textit{P}$_{3/2}$($\vert M_{J}\vert$=1/2) atoms from an intermediate 6\textit{S} state. In this case a single F$\ddot{\mathrm{o}}$rster resonance is observed at 1.79 V/cm [23]. Single resonance is most appropriate for the experimental and theoretical analysis and for implementing the quantum gates, because the overlapping of several resonances leads to decoherence. 

\begin{figure}
\includegraphics[scale=0.47]{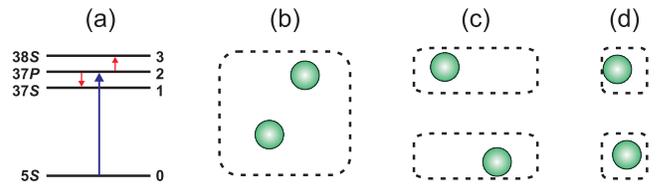}
\caption{\label{Fig1} (Color online) (a) Scheme of the relevant energy levels and transitions in Rb Rydberg atoms. Laser radiation initially excites the 37\textit{P} state from the 5\textit{S} ground state. Resonant dipole-dipole interaction (F$\ddot{\mathrm{o}}$rster resonance) induces transitions to the neighboring 37\textit{S} and 38\textit{S} states. (b)-(d) Typical spatial atom distributions in experiments with two interacting Rydberg atoms: (b) two atoms randomly placed in a single interaction volume, (c) two atoms in the individual cigar-shaped dipole traps, and (d) two atoms in the individual dipole traps evenly localized in all directions.}
\end{figure}

Our F$\ddot{\mathrm{o}}$rster resonance induces transitions between Rydberg states with $\Delta {M}_{{J}} =0$. This corresponds to the \textit{z}-oriented dipoles, and Eq.~\eqref{Eq1} reduces to a simpler form:

\begin{equation} \label{Eq2} 
\hat{V}_{ab} =\frac{\hat{d}_{a} \hat{d}_{b} }{4\pi \varepsilon _{0} } \left[\frac{1}{R_{ab}^{3} } -\frac{3\, \, Z_{ab}^{2} }{R_{ab}^{5} } \right],   
\end{equation} 

\noindent where $\hat{d}_{a,b} $ are the \textit{z} components of the dipole-moment operators of the two atoms and $Z_{ab} $ is the \textit{z} component of $R_{ab} $ (\textit{z} axis is chosen along the dc electric field).

For the short interaction time $t_{0} \le 1\; \mu {\rm s}$, which is of interest for fast quantum gates, we can ignore the spontaneous decay and blackbody-radiation-induced transitions from our Rydberg states (the effective lifetimes are tens of microseconds for $n>30$ [24]). This also simplifies the calculations by ignoring the motion of the atoms (displacement $<1\; \mu {\rm m}$ for atom temperatures $T<200\; \mu {\rm K}$) and the hyperfine structure ($<500\; {\rm kHz}$ for $n>30$). Then, in order to calculate the time evolution of the two Rydberg atoms, we can simply solve the Schr\"{o}dinger equation for a quasimolecule formed by two atoms. For the motionless atoms this problem is solved analytically. Let us denote the lower energy state 37\textit{S} as state ${\left| 1 \right\rangle} $, the middle state 37\textit{P}$_{3/2}$ as state ${\left| 2 \right\rangle} $, and the upper state 38\textit{S} as state ${\left| 3 \right\rangle} $, as shown in Fig.~1(a). The collective wave function for two Rydberg atoms is

\begin{equation} \label{Eq3} 
\Psi =a_{22} (t){\left| 22 \right\rangle} +a_{13} (t){\left| 13 \right\rangle} +a_{31} (t){\left| 31 \right\rangle} ,   
\end{equation} 

\noindent where ${\left| 22 \right\rangle} $ is the initial collective state populated by a short laser pulse at $t=0$, ${\left| 13 \right\rangle} $ and ${\left| 31 \right\rangle} $ are the two equally populated final states having a small energy detuning $\Delta$ from state ${\left| 22 \right\rangle} $, and $a_{ij} (t)$ are the time-dependent probability amplitudes to be found. The ground state 5\textit{S}, denoted as state ${\left| 0 \right\rangle} $ in Fig.~1(a), is ignored here as it does not contribute to the F$\ddot{\mathrm{o}}$rster resonance. We also ignore the other collective Rydberg states ${\left| 21 \right\rangle} ,{\left| 12 \right\rangle} ,{\left| 23 \right\rangle} ,{\left| 32 \right\rangle} $. These states have large energy detunings from state ${\left| 22 \right\rangle} $ and are not populated by DDI at the F$\ddot{\mathrm{o}}$rster resonance. A short laser pulse implies that the Fourier width of the laser radiation is large and it precludes the dipole blockade effect at laser excitation.

The signal measured in our experiments is a fraction of Rydberg atoms in final state ${\left| 1 \right\rangle} $ or a population of final state ${\left| 1 \right\rangle} $ per atom, which is calculated for the interaction time $t_{0} $ as

\begin{equation} \label{Eq4} 
\rho _{2} =\frac{1}{2} \left(\left|a_{13} (t_{0} )\right|^{2} +\left|a_{31} (t_{0} )\right|^{2} \right) .    
\end{equation} 

\noindent The solution of the Schr\"{o}dinger equation with the initial conditions $a_{22} (0)=1$ and $a_{13} (0)=a_{31} (0)=0$ then yields

\begin{equation} \label{Eq5} 
\rho _{2} = \frac{1}{2} \frac{2\Omega _{ab}^{2} }{2\Omega _{ab}^{2} +\Delta ^{2} /4} \sin ^{2} \left(\sqrt{2\Omega _{ab}^{2} +\Delta ^{2} /4} \; t_{0} \right),   
\end{equation} 

\noindent where $\Omega _{ab} =V_{ab} /\hbar $ is the DDI energy in the circular frequency units, and $\Delta =\left(2E_{2} -E_{1} -E_{3} \right)/\hbar $ is the detuning from exact F$\ddot{\mathrm{o}}$rster resonance, which is varied by the electric field. The interaction between two motionless Rydberg atoms thus leads to  population oscillations in the quasimolecule similarly to the Rabi oscillations in a two-level atom. Since this process is coherent, it can be used to implement two-qubit QPG. At $\Delta =0$ the first oscillation minimum in Eq.~\eqref{Eq5} corresponds to the $\pi $ phase shift of the collective wave function, which is necessary for QPG. Experimental observation and study of the Rabi-like oscillations is thus a prerequisite for quantum-information processing with neutral atoms.

\subsection{Two spatially fluctuating atoms}

In the realistic experiments the atom positions are not fixed even for the cold atoms in tightly focused dipole traps, due to finite atom temperatures [16]. The interaction strength is therefore a fluctuating value that may result in decoherence and washing out the Rabi-like oscillations in Eq.~\eqref{Eq5}. To calculate the signals measured in the experiments, Eq.~\eqref{Eq5} must be averaged over all possible spatial positions of the two interacting atoms. 

In the experiments with cold trapped atoms there are three spatial distributions that deserve special considerations. The first distribution is a completely disordered ensemble of the atoms confined in a small single excitation volume [Fig.~1(b)], which is formed by two intersecting laser beams, as in our experiment [23]. The second distribution is trapping of the individual atoms into close cigar-shaped dipole traps obtained at the focal points of the two laser beams [Fig.~1(c)], as in Refs. [9,10]. The third distribution is trapping of the individual atoms into close dipole traps evenly localized in all directions [Fig.~1(d)], like in two- or three-dimensional optical or magnetic lattices at low atom temperatures (see, e.g., Ref. [25]). These configurations appear to be most typical for the future experiments on quantum-information processing. We are interested in analyzing how their intrinsic atom position uncertainties would modify the spectrum of the ideal F$\ddot{\mathrm{o}}$rster resonance given by Eq.~\eqref{Eq5}.

No analytical solution can be obtained for the arbitrary spatial distributions. However, we have found that for a disordered ensemble of Fig.~1(b) it is possible to obtain a formula for the spatially averaged resonance amplitude $\langle\rho _{2} (\Delta =0)\rangle$. In order to do this we need to choose a proper spatial distribution of the atoms to define probability \textit{P}(\textit{r}) to find two atoms at distance \textit{r} in the excitation volume. The nearest-neighbor probability distribution by Chandrasekhar [26] is appropriate for two atoms:

\begin{equation} \label{Eq6} 
P(r)=e^{-r^{3} /r_{0}^{3} } 3r^{2} /r_{0}^{3}  ,  
\end{equation} 

\noindent where $r_{0} \approx \left[3/(4\pi n_0)\right]^{1/3} $ is the average distance between nearest-neighbor atoms at volume density $n_0$. It is valid for the atoms homogeneously distributed in the excitation volume. 

Another simplification we use is a conventional removal of the orientation-dependent part in Eq.~\eqref{Eq2}; that is, we consider a "scalar" dipole-dipole interaction. We remove the $3Z^2_{ab}/R^5_{ab} $ term in Eq.~\eqref{Eq2}, so that the interaction strength is now given by $1/R^3_{ab}$. Its validity for disordered atom ensembles is grounded by the fact that mutual atom orientations are effectively averaged in such ensembles. It will be tested later by comparing with numerical results. Then the resonance amplitude in Eq.~\eqref{Eq5} is averaged as 

\begin{equation} \label{Eq7} 
\langle\rho _{2} (\Delta =0)\rangle\approx \int _{0}^{\infty }\frac{1}{2} \sin ^{2}  \left(\frac{\sqrt{2} d_{21} d_{23} }{4\pi \varepsilon _{0} \hbar r^{3} } t_{0} \right)\; P(r)dr,   
\end{equation} 

\noindent where $d_{21}$ and $d_{23}$ are the dipole moments of transitions ${\left| 2 \right\rangle} \to {\left| 1 \right\rangle} $ and ${\left| 2 \right\rangle} \to {\left| 3 \right\rangle} $. For convenience, we replace the expression in the brackets with $\theta _{0} r_{0}^{3} /r^{3} $, where 

\begin{equation} \label{Eq8} 
\theta _{0} =\frac{\sqrt{2} d_{21} d_{23} }{4\pi \varepsilon _{0} \hbar r_{0}^{3} } t_{0}  
\end{equation} 

\noindent is the interaction pulse area at the average distance $r_{0} $. Further simplification of Eq.~\eqref{Eq7} is made by the replacement $x=r^{3} /r_{0}^{3} $ that yields

\begin{equation} \label{Eq9} 
\langle\rho _{2} (\Delta =0)\rangle\approx \int _{0}^{\infty }\frac{1}{2} \sin ^{2}  \left(\frac{\theta _{0} }{x} \right)\; e^{-x} dx.     
\end{equation} 

An approximate analytical solution may be found if we note that the main contribution to the integral in Eq.~\eqref{Eq9} comes from the very short distances, where rapidly oscillating squared sinus function can be replaced with its average value 1/2. Then the integration should start at $x=0$ and stop at some point $x_{c} \approx \alpha \theta _{0} $, where $\alpha \sim 1$, because the integrand exponentially drops at $x>x_{c} $. This approach finally gives a simple result:

\begin{equation} \label{Eq10} 
\langle\rho _{2} (\Delta =0)\rangle\approx \frac{1}{4} \left(1-e^{-\alpha \theta _{0} } \right).   
\end{equation} 

\noindent This is a general formula that describes both weak dipole-dipole interaction (amplitude grows as $\alpha \theta _{0} /4$) and strong interaction (amplitude saturates at 1/4). It may be used to calculate resonance amplitude for the arbitrary dipole moments, atom densities and interaction times, which are all included in $\theta _{0} $. The only fitting parameter is $\alpha $, which can be found by comparing Eq.~\eqref{Eq10} with more precise numerical calculations.

\subsection{Monte Carlo averaging for two disordered atoms}

Validity of Eq.~\eqref{Eq10} has been tested by numerical Monte Carlo simulations. In this approach, the two Rydberg atoms are randomly placed in a single interaction volume as in Fig.~1(b), and the solution given by Eq.~\eqref{Eq5} is averaged over many random position realizations ($\sim 10^{4} $). 

The model interaction volume is chosen to be a cube of size \textit{L}, in order to assign random atom coordinates along the three Cartesian axes independently. This significantly simplifies the calculations, otherwise it is necessary to account for the possible correlations between the coordinates. Although actual excitation volume differs from a cube, special comparison between Monte Carlo simulations for the spherical and cubic volumes has shown that the results are almost identical if the excitation volumes are equal. Therefore, in the further calculations we use only cubic or parallelepiped volumes with homogeneously distributed atoms. 

For the correct comparison with Eq.~\eqref{Eq10} we should properly define $r_{0} $ for two Rydberg atoms in a cubic volume $V=L^{3} $. A natural way is to make this connection via the volume density of the two atoms in a cubic volume $n_0=2/L^{3}=(4\pi r_{0}^{3} /3)^{-1} $ that yields $r_{0} \approx 0.5L$.

Figure 2 shows the numerical and analytical results obtained for the scalar (a) and full (b) DDI between two Rydberg atoms randomly placed in a cubic interaction volume. From Fig.~2(a) we see that usage of the scalar approximation results in the resonance amplitude that exceeds the saturated value 1/4 at $\theta _{0} \approx 2-14$. The reason is a residual coherence due to incompletely damped Rabi oscillations, which are not taken into account in Eq.~\eqref{Eq10}. Exact numerical solution of Eq.~\eqref{Eq9} reproduces well this peculiarity. This effect is completely absent in Fig.~2(b) where Monte Carlo simulation is done for the full DDI. Random orientation of the dipoles washes out the Rabi-like oscillations at any $\theta _{0} $, so that no coherence is expected for the F$\ddot{\mathrm{o}}$rster resonance between two disordered atoms.

A very surprising finding is that Eq.~\eqref{Eq10} fits well the numerical simulations in Fig.~2(b) at $\alpha =0.55$, although this equation has been derived for the scalar DDI. We suggest that Eq.~\eqref{Eq10} is a universal formula that can be used to obtain the spatially averaged F$\ddot{\mathrm{o}}$rster resonance amplitude in disordered atom ensembles. For example, a similar formula was used empirically in Ref. [14] to fit the time dependence of the number of interacting Rydberg atoms excited by a laser pulse. Universal scaling in a strongly interacting Rydberg gas was discussed in Ref.~[27].

\begin{figure}
\includegraphics[scale=0.45]{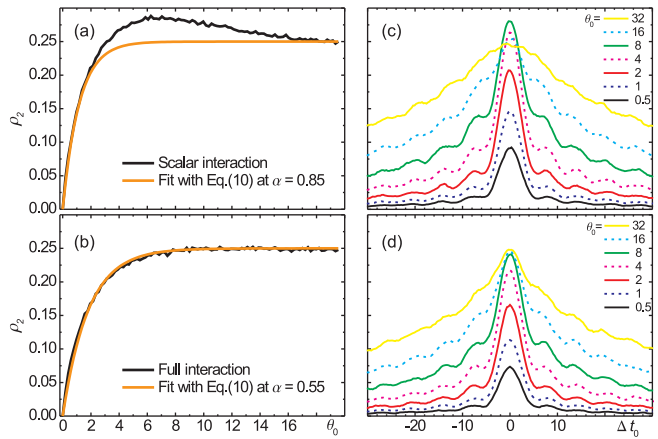}
\caption{\label{Fig2}(Color online) Monte Carlo simulation of a F$\ddot{\mathrm{o}}$rster resonance between two Rydberg atoms randomly placed in a cubic interaction volume [as in Fig.~1(b)]. (a) and (b) Dependence of the spatially averaged resonance amplitude on the average interaction pulse area $\theta _{0} $ for the scalar (a) and  full (b) dipole-dipole interaction; analytical fit has been done with Eq.~\eqref{Eq10}. (c) and (d) Spectra of the F$\ddot{\mathrm{o}}$rster resonance in the dimensionless detuning scale $\Delta \: t_{0} $ at various $\theta _{0} $ for the scalar (c) and  full (d) dipole-dipole interaction.}
\end{figure}

In Fig.~2 we also present the numerically calculated spectra of the F$\ddot{\mathrm{o}}$rster resonance in the dimensionless detuning scale $\Delta \: t_{0} $ for the scalar (c) and full (d) DDI obtained at various $\theta _{0} $, that is, at various interaction volumes or interaction times. These spectra clearly show how the resonance saturates and broadens as $\theta _{0} $ grows. Below saturation ($\theta _{0} \le 1$) the ultimate full width at half maximum (FWHM) is equal to $2\pi $, corresponding to $2\pi /t_{0} $ in the circular frequency scale. It is mainly a Fourier-transform limited width, which is larger than the average energy of DDI given by $\theta _{0} /t_{0} $. For $\theta _{0} >1$ the resonance broadens and its amplitude saturates at the 0.25 value. Averaging over the atom positions forms a resonance with broad wings and a cusp on the top, which is similar to that observed in atomic beams [28,29]. 

As we pointed out in Ref.~[23], the resonance line shapes cannot be properly fit with a Lorentz profile. When we match the central part, the resonance wings are much broader than the Lorentz ones, due to rare interactions at very short distances. The commonly used Lorentz fit is thus inadequate for precise comparison between theory and experiment. 

Although the residuals of the Rabi-like population oscillations remain in the spectra, from Fig.~2 we see that DDI in the disordered atom ensemble is incoherent and cannot be used to implement quantum phase gates. The two atoms must be spatially localized in the individual dipole traps to improve the coherence.

\subsection{Monte Carlo averaging for two localized atoms}

If the two atoms are localized in the individual dipole traps as in Figs.~1(c)-1(d), the interaction strength depends on the spatial orientation of the traps with respect to each other and to the quantization axis \textit{z}. The quantization axis is most convenient to choose along the dc electric field applied for Stark tuning of the F$\ddot{\mathrm{o}}$rster resonance and for field-ionization detection of Rydberg atoms. According to Eq.~\eqref{Eq2}, the strongest dipole-dipole interaction occurs when the two atoms are aligned along the \textit{z} axis. In this case the axes of the two dipole traps in Fig.~1(c) are directed orthogonally to \textit{z}, and spatial fluctuations occur mainly in the $x-y$ plane. We are free to choose the dipole trap axes along the \textit{x} direction for distinctness. In what follows we shall consider only this spatial configuration for the two atoms localized in the identical individual traps. 

Let us denote $r_{x} ,\; r_{y}$, and $r_{z} $ to be the spatial uncertainties of the two atoms in the $x,\; y$, and $z$ directions, correspondingly, and $R_{0} $ to be the distance between the trap axes. The average interaction pulse area is defined for this configuration as

\begin{equation} \label{Eq11} 
\theta _{0} =\frac{2\sqrt{2} d_{21} d_{23} }{4\pi \varepsilon _{0} \hbar R_{0}^{3} } t_{0} .    
\end{equation} 

We also need to choose the realistic spatial parameters of the dipole traps. Experiment [10] can be used for reference. The distance between the two traps was $R_{0} \approx 10\; \mu {\rm m}$. At the atom temperatures of $\sim 200\; \mu {\rm K}$, the reported position probability distributions for each atom were approximately Gaussian with $\sigma _{y,z} \sim 0.3\; \mu {\rm m}$ and $\sigma _{x} \sim 4\; \mu {\rm m}$, corresponding to FWHM of $r_{y,z} \approx 0.7\; \mu {\rm m}$ and $r_{x} \approx 9.4\; \mu {\rm m}$. This dipole trap geometry can also be characterized by the aspect ratio $r_{x} /r_{y,z} \approx 13.4$. 

\begin{figure}
\includegraphics[scale=0.5]{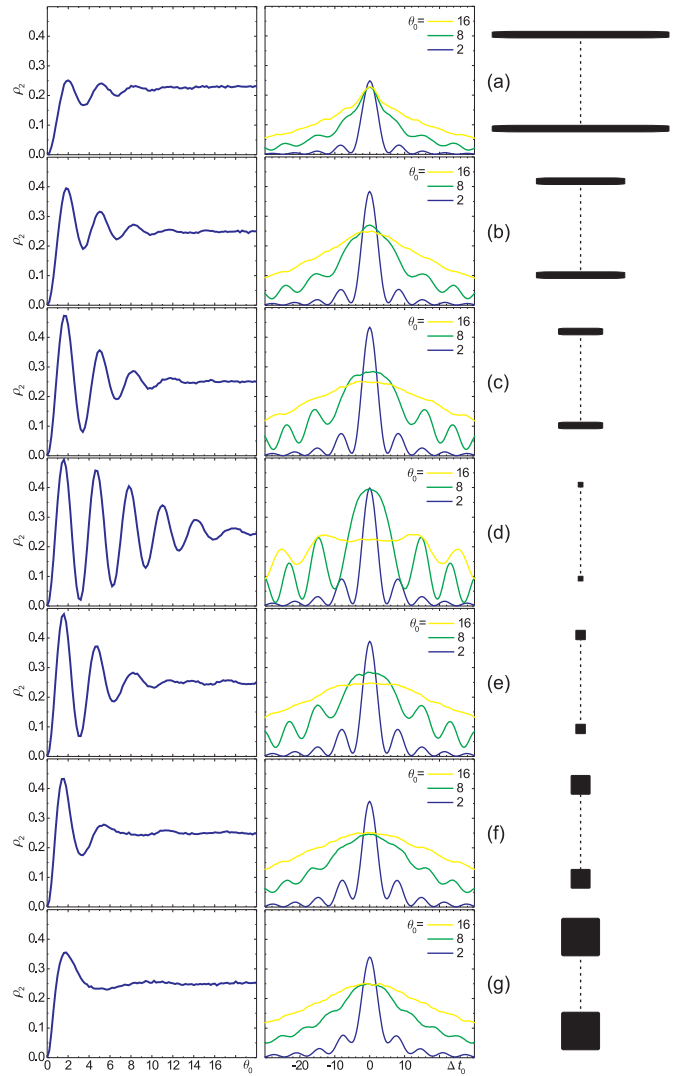}
\caption{\label{Fig3} (Color online) Monte Carlo simulation of a F$\ddot{\mathrm{o}}$rster resonance between two Rydberg atoms randomly placed in the two individual cigar-shaped [(a)-(c)] or symmetrical [(d)-(g)] dipole traps of various size. Left-hand panels: Dependence of the spatially averaged resonance amplitude on the average interaction pulse area $\theta _{0} $. Middle panels: Spectra of the F$\ddot{\mathrm{o}}$rster resonance in the dimensionless detuning scale $\Delta \: t_{0} $ for various $\theta _{0} $. The right-hand pictures show to scale the relative atom position uncertainties and the distance between the traps. The \textit{z} axis is oriented vertically.}
\end{figure}

Figures 3(a)-3(c) present our numerical Monte Carlo results obtained for the dipole trap geometry described previously, but the aspect ratio is varied from (a) 26.8 to (b) 13.4 and (c) 6.7. The right-hand pictures in Fig.~3 show to scale the relative atom position uncertainties and the distance between the traps. Dependencies of the resonance amplitude on $\theta _{0} $ (the left-hand panels in Fig.~3) demonstrate the damped Rabi-like population oscillations. The first oscillation minimum around $\theta _{0} \approx \pi$ corresponds to the $\pi $ phase shift of the collective wave function. This point is what is needed to realize QPG. Nonzero minimum at this point results in a finite fidelity of QPG. Fidelity is given by the probability of the two atoms to remain in the initial state $\vert 22 \rangle$ after this collective state obtains the phase shift $\pi$: $F=1-2 \rho_2(\theta_0=\pi)$.

From Fig.~3 it is seen that fidelity ($F\approx0.62$) is poor for the trap geometry of Fig.~3(b) corresponding to the experimental conditions of Ref.~[10]. An acceptable QPG fidelity of 0.85 can be obtained for the geometry of Fig.~3(c), that is, when the atoms are localized along the \textit{x} axis two times better. This requires the atom temperature to be decreased from $200\; \mu {\rm K}$ to $\sim 50\; \mu {\rm K}$. The related spectra in Fig.~3 (the middle panels) show that coherence is lost as $\theta _{0} $ increases. One may conclude that the coherent dipole-dipole interaction of Rydberg atoms and its usage in QPG require the atoms to be colder and localized better than in the state-of-the-art dipole traps. 

We therefore have done the calculations for the symmetrical dipole traps evenly localized in all directions [Figs.~3(d)-3(g)], which apparently can be realized at the lower atom temperatures. The highest coherence and fidelity of 0.96 are observed in Fig.~3(d) where $r_{x,y,z} =R_{0} /20$. Such geometry corresponds to the 0.5~$\mu {\rm m}$ in size dipole traps separated by 10~$\mu {\rm m}$.  This is close to the ultimate spatial resolution achievable with optical dipole traps. A somewhat worse fidelity of 0.86 is observed in Fig.~3(e) for the more realistic 1~$\mu {\rm m}$ in size dipole traps. Larger traps of 2~$\mu {\rm m}$ (f) and 4~$\mu {\rm m}$ (g) in size demonstrate stronger decoherence (0.66 and 0.53 fidelities, correspondingly) and are not suitable for QPG. Overall, Fig.~3 helps to estimate the attainable coherence and fidelity of QPG for various trap configurations.

\subsection{Monte Carlo averaging for dipole blockade between two atoms}

Dipole blockade (DB) must appear as the laser excitation of only one Rydberg atom out of a ground-state mesoscopic ensemble if the Rydberg atoms are strongly interacting [2]. It has been observed recently for just two Rydberg atoms interacting via quasi-F$\ddot{\mathrm{o}}$rster resonances in two close dipole traps [9,10]. 

In this section we will simulate DB at the Stark-tuned F$\ddot{\mathrm{o}}$rster resonance between two Rydberg atoms. There have been several articles on the numerical modeling of DB for many atoms (see, e.g., [30-37]). Most articles analyzed the excitation probabilities in the resonance center, and only a few of them calculated the DB laser excitation spectra for some particular cases [31]. Here we present the extended calculations of the laser excitation spectra for two atoms in various spatial configurations, and analyze the effects of the F$\ddot{\mathrm{o}}$rster resonance detuning and laser non-monochromaticity.

In order to calculate the dipole blockade spectrum for two atoms, we shall consider the narrowband laser excitation of Rydberg state ${\left| 2 \right\rangle} $ from ground state ${\left| 0 \right\rangle} $ [see Fig.~1(a)] for the two interacting atoms in the quasimolecular basis: 

\begin{equation} \label{Eq12} 
\begin{array}{c} {\Psi =a_{00} (t)|00\rangle+a_{20} (t)|20\rangle+a_{02} (t)|02\rangle+} \\ {a_{22} (t)|22\rangle+a_{13} (t)|13\rangle+a_{31} (t)|31\rangle .} \end{array}  
\end{equation} 

\noindent Other collective states have large energy detunings and are not populated by the exciting laser radiation or DDI. The Hamiltonian is now taken as

\begin{equation} \label{Eq13} 
\hat{H}=\left(\hat{d}_{a} +\hat{d}_{b} \right)E(t)+\frac{\hat{d}_{a} \hat{d}_{b} }{4\pi \varepsilon _{0} } \left[\frac{1}{R_{ab}^{3} } -\frac{3\, \, Z_{ab}^{2} }{R_{ab}^{5} } \right],  
\end{equation} 

\noindent where $E(t)$ is the electric field of the laser radiation. Laser-induced transitions occur from initial ground state  ${\left| 00 \right\rangle} $ to single-excited states $|20\rangle$ and $|02\rangle$, and to double-excited state $|22\rangle$. State ${\left| 22 \right\rangle} $ is coupled to other double-excited states ${\left| 13 \right\rangle} $ and ${\left| 31 \right\rangle} $, via DDI. Strong interaction shifts state ${\left| 22 \right\rangle} $ out of the resonance with laser radiation, and dipole blockade results in the decrease of the excitation probability of double-excited states.

Probabilities to excite one or two atoms by a square laser pulse of duration $t_{0} $ are calculated as 

\begin{equation} \label{Eq14} 
\begin{array}{l} {P_{1} =\left|a_{20} (t_{0} )\right|^{2} +\left|a_{02} (t_{0} )\right|^{2}, } \\ {P_{2} =\left|a_{22} (t_{0} )\right|^{2} +\left|a_{31} (t_{0} )\right|^{2} +\left|a_{13} (t_{0} )\right|^{2} .} \end{array}     
\end{equation} 

\noindent Blockade fidelity is given by $P_{1} $. Perfect blockade corresponds to $P_{1} =1$ and $P_{2} =0$.

In our Monte Carlo calculations the laser radiation is assumed to be linearly polarized along the \textit{z} axis. Laser detuning $\delta $ from the exact resonance ${\left| 0 \right\rangle} \to {\left| 2 \right\rangle} $ is varied around its unperturbed zero value. Monte Carlo excitation spectra are calculated in the dimensionless detuning scale $\delta \: t_{0} $. The laser excitation pulse area  is $\Theta =d_{02} E_{0} t_{0} /\hbar $, where $d_{02} $ is the dipole moment of transition ${\left| 0 \right\rangle} \to {\left| 2 \right\rangle} $ and $E_{0} $ is the electric field amplitude of the laser radiation. If  $\Theta =\pi $ and $\delta =0$, the non-interacting atoms are completely transferred to Rydberg state ${\left| 2 \right\rangle} $. In the interacting atoms the probability of transfer to double-excited states reduces due to DB.

In contrast to Eq.~\eqref{Eq5}, there is no simple analytical solution to the Schr\"{o}dinger equation for dipole blockade. Therefore, all further simulations have been done numerically. Figure 4 shows the calculated Monte Carlo spectra of the one- and two-atom laser excitation probabilities $P_{1}$ and $P_{2} $ for $\Theta =\pi $ and for various average interaction pulse areas $\theta _{0} $. Two Rydberg atoms are randomly placed in the two individual cigar-shaped [Figs.~4(a)-4(c)] or symmetrical [Figs.~4(d)-4(g)] dipole traps of various size or in a single excitation volume [Fig.~4(h)] [for the latter configuration $\theta _{0} $ is defined according to Eq.~\eqref{Eq8}]. The right-hand pictures show to scale the relative atom position uncertainties and the distance between the traps, which are the same as in Fig.~3. F$\ddot{\mathrm{o}}$rster resonance detuning $\Delta $ is taken to be zero.

\begin{figure}
\includegraphics[scale=0.47]{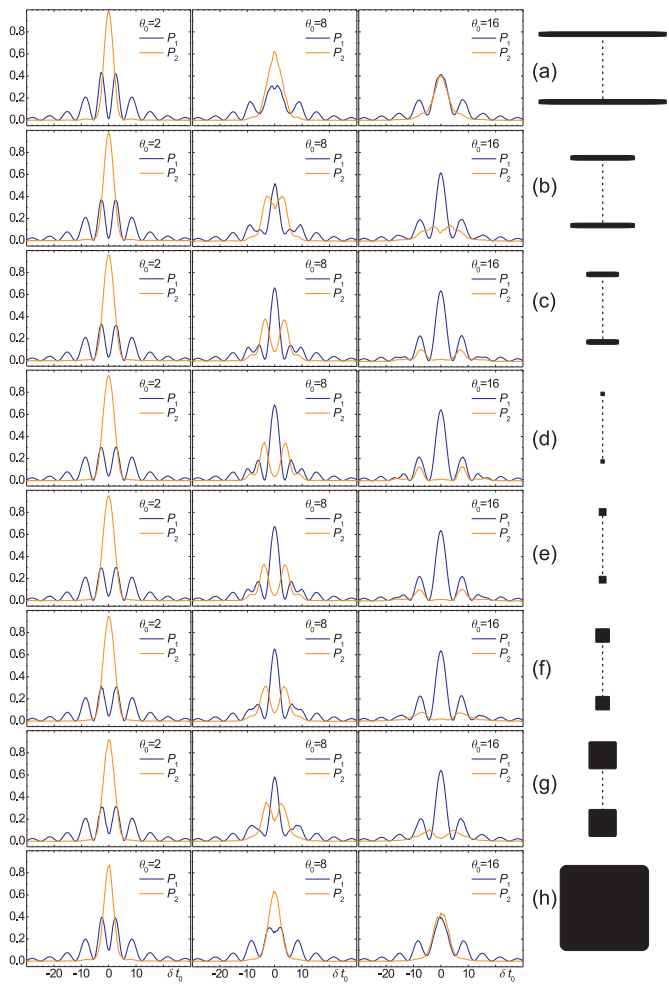}
\caption{\label{Fig4} (Color online) Monte Carlo simulation of the dipole blockade at a F$\ddot{\mathrm{o}}$rster resonance between two Rydberg atoms randomly placed in two individual cigar-shaped [(a)-(c)] or symmetrical [(d)-(g)] dipole traps of various sizes and in a single excitation volume (h). Spectra of the one-atom and two-atom laser excitation probabilities, $P_{1}$ and $P_{2} $, are shown in the dimensionless detuning scale $\delta \: t_{0} $ for various average interaction pulse areas $\theta _{0} $. The right-hand pictures show to scale the relative atom position uncertainties and the distance between the traps. The \textit{z} axis is oriented vertically. The laser excitation pulse area is $\Theta =\pi $. The F$\ddot{\mathrm{o}}$rster resonance detuning $\Delta $ is zero.}
\end{figure}

From the left-hand panels in Fig.~4 it is seen that there is almost no blockade at $\theta _{0} =2$ for any trap configuration.  These spectra are just probabilities to populate one- and two-atom states for the two non-interacting atoms, which are identical to what we observed experimentally at microwave transitions between sodium Rydberg states in our earlier work that relied on the atom-number resolved study [15]. In the line center $(\delta =0)$ probability to excite both atoms, $P_{2} $, is close to unity, while probability to excite one of the two atoms, $P_{1} $, has a dip and is nearly zero. Rabi oscillations appear in the resonance wings for our square laser excitation pulse.

Dipole blockade becomes apparent at $\theta _{0} =8$ (middle panels in Fig.~4). Probability $P_{2} $ reduces in the line center and there appears a dip corresponding to either partial [(a), (b), (g), (h)] or almost complete [(c)-(f)] dipole blockade. At the same time, $P_{1} $ grows and its dip converts to a peak. At $\theta _{0} =16$ (right-hand panels in Fig.~4), these features transform to a strong blockade effect in the line center. In Figs.~4(c)-4(f) probability $P_{2} $ drops to $\approx 0.013-0.024$, while $P_{1} $ saturates at $\approx 0.64$. 

Large atom position uncertainties in Figs.~4(a) and 4(h) substantially reduce the blockade efficiency at $\theta _{0} =16$ due to zeros of the F$\ddot{\mathrm{o}}$rster interactions at certain dipole orientations [8] and due to lower interaction energy between our \textit{z}-dipoles in the $x-y$ plane. Nevertheless, we have found that blockade is still achievable for these spatial configurations. Blockade fidelity of 95\% is reached at $\theta _{0} >100$. This requires the interaction distance to be $\approx 2-2.5$ times smaller than that for the well localized atoms in Figs.~4(c)-4(f).

Further analysis is done with Fig.~5. The left-hand panels in Fig.~5 demonstrate that the $P_{1} $ value can be increased up to 1 if we set a smaller laser pulse area $\Theta =\pi /\sqrt{2} $. This $\sqrt{2} $ reduction is due to the $\sqrt{N} $ dependence of the collective Rabi frequency on the number of interacting atoms \textit{N} [30]. In Fig.~5 we use this new $\Theta $ value in all simulations. 

The middle panels in Fig.~5 show the dependencies of the blockade fidelity ($P_{1} $ value) on the average dipole interaction pulse area $\theta _{0} $ for three different detunings from the exact F$\ddot{\mathrm{o}}$rster resonance: $\Delta =0,\; 20\pi /t_{0}$, and $40\pi /t_{0} $. These correspond to $\Delta /2\pi =$0, 10, and 20 MHz detunings at the $t_{0} =1\; \mu {\rm s}$ interaction time. Zero detuning is relevant to the precisely Stark-tuned F$\ddot{\mathrm{o}}$rster resonances, while 5-20 MHz detunings are typical for accidental quasi-F$\ddot{\mathrm{o}}$rster resonances [8]. It is seen that dipole blockade has a threshold-like behavior versus $\theta _{0} $, and non-zero detuning substantially increases the threshold. Nevertheless, the detuning effect can be overcome by a stronger interaction, so that it is still possible to reach $P_{1} \approx 1$. This has been confirmed experimentally in Refs. [9,10].

Figures 4 and 5 help us to understand what kind of one- and two-atom excitation spectra can be observed in the presence of DB. We note, however, that these calculations were made for a monochromatic laser radiation (the laser linewidth $\Delta \omega _{L} \ll 2\pi /t_{0} $) and square laser excitation pulse, which produce the Rabi-like oscillations in the spectra. The oscillations are washed out if $\Delta \omega _{L} $ increases due to technical imperfections or if parasitic electromagnetic fields broaden the resonance. This leads to decoherence and decreases the blockade fidelity. The right-hand panels in Fig.~5 show the dependencies of $P_{1} $ on $\theta _{0} $ for the three different laser linewidths: $\Delta \omega _{L} =0,\; 2\pi /t_{0}$, and $4\pi /t_{0} $. These correspond to $\Delta \omega _{L} /2\pi =$0, 1, and 2 MHz at the $t_{0} =1\; \mu {\rm s}$ interaction time. Finite laser linewidth in our Monte Carlo simulation was introduced as random laser frequency jumps around the zero value. It is seen that laser line width inevitably decreases the dipole blockade fidelity, which cannot be improved by a stronger interaction. Therefore, depending on the linewidth of the available laser, it is crucial to use $t_{0} \ll 2\pi /\Delta \omega _{L} $ in the experiments in order to achieve high blockade and quantum gate fidelities.

\begin{figure}
\includegraphics[scale=0.47]{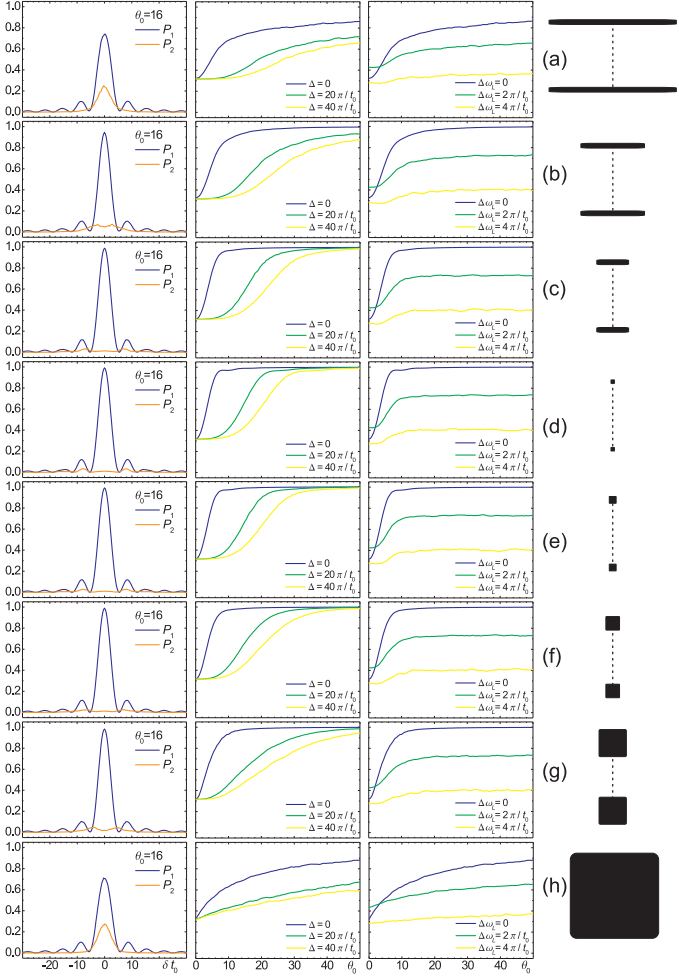}
\caption{\label{Fig5} (Color online) Monte Carlo simulation of the dipole blockade at a F$\ddot{\mathrm{o}}$rster resonance between two Rydberg atoms randomly placed in two individual cigar-shaped [(a)-(c)] or symmetrical [(d)-(g)] dipole traps of various sizes and in a single excitation volume (h). The left-hand panels demonstrate the dipole blockade spectra at the laser pulse area $\Theta =\pi /\sqrt{2} $. The middle panels show the dependencies of the blockade fidelity $P_{1} $ on the average dipole interaction pulse area $\theta _{0} $ for three different detunings from the exact F$\ddot{\mathrm{o}}$rster resonance: $\Delta =0,\; 20\pi /t_{0}$, and $40\pi /t_{0} $. The right-hand panels show the dependencies of $P_{1} $ on $\theta _{0} $ for three different laser linewidths: $\Delta \omega _{L} =0,\; 2\pi /t_{0}$, and $4\pi /t_{0} $. The right-hand pictures show to scale the relative atom position uncertainties and the distance between the traps.}
\end{figure}

\section{INTERACTION OF MORE THAN TWO RYDBERG ATOMS}

\subsection{Frozen atoms}

Let us now consider a F$\ddot{\mathrm{o}}$rster resonance between $N>2$ frozen Rydberg atoms with ${\left| 1 \right\rangle} $, ${\left| 2 \right\rangle} $, and ${\left| 3 \right\rangle} $ Rydberg levels participating in the resonance [see Fig.~1(a)]. The main calculation difficulty is that basis size $Z(N)$ of the collective wave function rapidly grows with \textit{N}. It is equal to the Gegenbauer polynomial $C_{N}^{(-N)} (1/2)$, which yields $Z(2)=3$, $Z(3)=7$, $Z(4)=19$, $Z(5)=51$, $Z(6)=141$, ... , $Z(10)=8953$, ... . 

We assume that at $t=0$ a short laser pulse initially excites the collective state ${\left| 22222...2 \right\rangle} $. This state is coupled via F$\ddot{\mathrm{o}}$rster resonance to $N!/(N-2)!$ states of the kind ${\left| 13222...2 \right\rangle} $. In turn, these states are coupled to $N!/[(N-4)!(2!)^{2} ]$ states of the kind ${\left| 13132...2 \right\rangle} $, and so on. Taking into account all possible permutations, a general formula for the number of such states is $n_{k} =N!/[(N-k)!((k/2)!)^{2} ]$, where \textit{k} is the number of Rydberg atoms that have abandoned state ${\left| 2 \right\rangle} $ due to DDI (\textit{k} is always even for the F$\ddot{\mathrm{o}}$rster resonance, because one atom goes to the lower Rydberg state and the other atom simultaneously goes to the upper state). The transition probability per atom, or fraction of atoms in final state ${\left| 1 \right\rangle} $, is calculated as

\begin{equation} \label{Eq15} 
\rho _{N} =\sum _{k=2,4,6...}\frac{k}{2N} \sum _{i=1}^{n_{k} }\left|a_{i\; k} \right|^{2}   ,   
\end{equation} 

\noindent where $a_{i\; k} $ are the probability amplitudes of collective states, and $k/(2N)$ is a normalization factor for each \textit{k}. The maximum \textit{k} value is equal to \textit{N} for the even \textit{N}, or to $N-1$ for the odd \textit{N}.

It is impossible to derive a simple analytical formula like Eq.~\eqref{Eq5} for $\rho _{N} $ at $N>2$, even for frozen Rydberg atoms. This is due to the fact that besides the resonant dipole-dipole interaction ${\rm Rb}(37P)+{\rm Rb}(37P)\to {\rm Rb}(37S)+{\rm Rb}(38S)$, it is necessary to take into account the exchange processes like ${\rm Rb}(37P)+{\rm Rb}(37S)\to {\rm Rb}(37S)+{\rm Rb}(37P)$ and ${\rm Rb}(37P)+{\rm Rb}(38S)\to {\rm Rb}(38S)+{\rm Rb}(37P)$, which are always resonant (independent of the electric field) and lead to a population hopping between neighboring atoms [17]. This hopping broadens and decoheres the F$\ddot{\mathrm{o}}$rster resonances, although it is not the main source of the broadening [38].

If, however, we consider a weak dipole-dipole interaction, most of the population remains in the initial state ${\left| 22222...2 \right\rangle} $, and only states with $k=2$ are weakly populated. In this case the always-resonant processes can be neglected, since\textit{ S} states in each atom are weakly populated too. Then it can be shown that Eq.~\eqref{Eq5} is simply modified by replacing $2\Omega _{ab}^{2} $ with the sum of the squares of the interaction energies for all atomic pairs: 

\begin{equation} \label{Eq16} 
\Omega ^{2} =\sum _{a\ne b}\Omega _{ab}^{2}    ,     
\end{equation} 

\noindent and Eq.~\eqref{Eq15} finally yields

\begin{equation} \label{Eq17} 
\rho _{N} \approx \frac{1}{N} \frac{\Omega ^{2} }{\Omega ^{2} +\Delta ^{2} /4} \sin ^{2} \left(\sqrt{\Omega ^{2} +\Delta ^{2} /4} \; t_{0} \right).  
\end{equation} 

\noindent Exact numerical modeling with the exchange interactions fully taken into account has confirmed that Eq.~\eqref{Eq17} works well at $\rho _{N} (t_{0} )<0.1$, both in the resonance center and wings. 

For the strong DDI the maximum $\rho _{N} $ value depends on the \textit{N} evenness. It is 1/2 for the even \textit{N} and $(N-1)/(2N)$ for the odd \textit{N}. In particular, for three atoms the maximum transfer probability is 1/3, since only one of the three atoms can be transferred to state ${\left| 1 \right\rangle} $ at a time. Decoherence due to spatial fluctuations may decrease these maximum attainable values by up to 2 times.

\subsection{Spatially fluctuating atoms}

We can imagine various spatial configurations including linear, planar, cubic, honeycomb, or ring lattices of single ground-state atoms intended for a quantum computation. Corresponding considerations were carried out in many theoretical papers (see, e.g., [30,32,33]).

Two-qubit logic gates imply simultaneous laser excitation of only two Rydberg atoms. Our results obtained in the previous section can be applied for their analysis with minor changes concerning mainly the renormalization of the interaction strength.

In this section we consider simultaneous laser excitation and interaction of more than two Rydberg atoms.  One may expect that many-body phenomena would enhance the DDI strength in the mesoscopic ensembles, but coherence may be lost upon averaging over many interacting atoms. 

The numerical Monte Carlo simulations were performed for $N=2-5$ interacting Rydberg atoms, initially excited to state  ${\left| 2 \right\rangle} $ and randomly positioned in a single cubic or cigar-shaped volume or in the evenly spaced individual dipole traps (see the right-hand pictures in Fig.~6). In this approach, the time evolution of all $Z(N)$ collective states is obtained by numerically solving the Schr\"{o}dinger equation for various \textit{N}, $\Delta $, and interaction times $t_{0} $. We accounted for all possible binary resonant interactions between \textit{N} atoms, as well as the always-resonant exchange interactions. The initial positions of \textit{N} atoms were averaged over 500 random realizations. A similar approach was used in Refs. [21,22].

Figure 6(a) shows the results of our calculations of the F$\ddot{\mathrm{o}}$rster resonance amplitude and spectrum for two to five Rydberg atoms randomly positioned in a cubic volume. The dependencies of the amplitude on the average interaction pulse area $\theta _{0} $ [defined for two atoms according to Eq.~\eqref{Eq8}] demonstrate similar saturation curves for all \textit{N}. The curves for \textit{N}=2, 4, and 5 saturate at $\rho _{N} =0.25$ and are fitted well with Eq.~\eqref{Eq10} at $\alpha_2 \approx 0.55$, $\alpha_4 \approx 1.60$, and $\alpha_5 \approx 2.10$. 

The case $N=3$ is distinguished by a different saturation amplitude $\rho _{3} \approx 0.21$ and a different saturation curve, which is similar to that in Fig.~2(a). The saturated amplitude lies between the maximum possible value of 1/3 for coherent interaction and 1/6 for completely incoherent interaction. This indicates that residual coherence is always present in the interaction of three atoms. In general, this observation should be typical for the odd number of the interacting Rydberg atoms. For example, the resonance amplitude for five atoms in Fig.~6(a) saturates at $\rho _{5} \approx 0.25$ and also lies between the maximum possible value of 2/5 for coherent interaction and 1/5 for incoherent interaction.

The spectra in Fig.~6(a) are calculated for the unsaturated ($\theta _{0} =0.5$) and saturated ($\theta _{0} =8$) F$\ddot{\mathrm{o}}$rster resonances. In the unsaturated spectra the ultimate FWHM is defined by the inverse interaction time $2\pi /t_{0} $ for all \textit{N}. It is mainly a Fourier-transform-limited width, unless $\alpha _{N} \theta _{0} <1$. The probability for two atoms to interact at shorter distances (where $\alpha _{N} \theta _{0} >1$) is low, and the resonance is narrow despite the spatial averaging. The signatures of the Rabi-like oscillations are visible in the resonance wings. The resonances broaden as \textit{N} grows, due to the increase in the average energy of DDI. 

\begin{figure}
\includegraphics[scale=0.47]{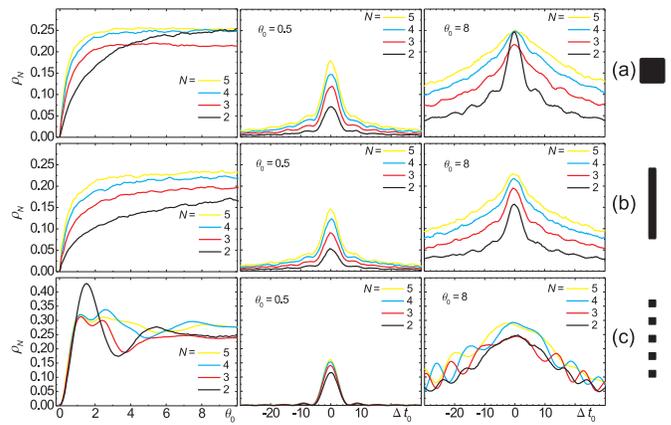}
\caption{\label{Fig6} (Color online) Monte Carlo simulation of the F$\ddot{\mathrm{o}}$rster resonance between $N=2-5$ Rydberg atoms randomly placed in (a) a single cubic excitation volume, (b) a single cigar-shaped dipole trap, and (c) individual evenly spaced symmetrical dipole traps. Left-hand panels: Dependence of the spatially averaged resonance amplitude on the average interaction pulse area between two atoms $\theta _{0} $. Middle panels: Spectra of the F$\ddot{\mathrm{o}}$rster resonance in the dimensionless detuning scale $\Delta \: t_{0} $ at $\theta _{0} =0.5$ and  $\theta _{0} =8$. The right-hand pictures show to scale the relative atom position uncertainties. The \textit{z} axis is oriented vertically.}
\end{figure}

Figure 6(b) presents the amplitude and spectra calculated for two to five atoms confined in a single cigar-shaped dipole trap oriented along the \textit{z} axis. The aspect ratio of the trap is $r_{z} /r_{x,y} =10$ (for example, a 10-$\mu {\rm m}$-long dipole trap with 1-$\mu {\rm m}$ transverse size, as in Ref.~[10]). As it was pointed out in Ref. [39], mainly nearest-neighbor atoms interact in this specific linear geometry. This reduces the saturated resonance amplitude to a value below 0.25, because for \textit{N} atoms in the elongated dipole trap there are less than \textit{N} effectively interacting atoms on the average.

Figure 6(c) shows the amplitude and spectra calculated for the linear dipole trap arrays of $N=2-5$ atoms aligned along the \textit{z} axis. The dipole traps are evenly spaced at distance \textit{r}, while the atom position uncertainties are \textit{r}/5 [for two atoms this geometry corresponds to Fig.~3(f)]. The value of $\theta _{0} $ is defined for two neighboring atoms according to Eq.~\eqref{Eq11}. As the atoms are localized well in all directions, the interaction is coherent to some extent. The maximum coherence is observed for two atoms. At $\theta _{0} =0.5$ the F$\ddot{\mathrm{o}}$rster resonance spectra are almost identical for all \textit{N} due to primarily nearest-neighbor interaction. At $\theta _{0} =8$ there appear the asymmetries in the spectra which have different characters for different \textit{N}.

Figure 6 demonstrates that in all spatial configurations coherence is lost as \textit{N }increases, although the DDI strength substantially grows. We therefore cannot benefit from increasing the number of the interacting Rydberg atoms in performing the quantum phase gates.

On the other hand, dipole blockade is less sensitive to decoherence and it might benefit from increasing \textit{N}. With our computer code we are able to calculate the DB excitation spectra and fidelity for up to five atoms in various spatial configurations. 

As an example, Fig.~7 presents the Monte Carlo results for the dipole blockade at a F$\ddot{\mathrm{o}}$rster resonance between three Rydberg atoms randomly placed in a single cubic excitation volume. Spectra of the one-, two-, and three-atom laser excitation probabilities $P_{1} ,\; P_{2}$, and $P_{3} $ are shown in the dimensionless detuning scale $\delta \: t_{0} $ for various average interaction pulse areas $\theta _{0} $ [defined for two atoms according to Eq.~\eqref{Eq8}]. The laser excitation pulse area is $\Theta =\pi $ in Figs.~7(a)-7(c) or $\Theta =\pi /\sqrt{3} $ in Fig.~7(d). The F$\ddot{\mathrm{o}}$rster resonance detuning $\Delta $ is zero.

\begin{figure}
\includegraphics[scale=0.5]{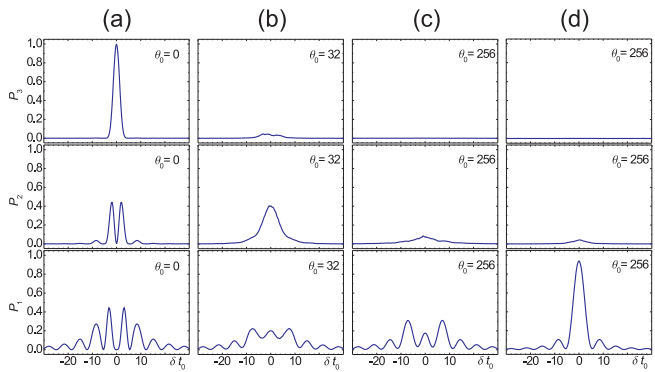}
\caption{\label{Fig7} (Color online) Monte Carlo simulation of the dipole blockade at a F$\ddot{\mathrm{o}}$rster resonance between three Rydberg atoms randomly placed in a single cubic excitation volume. Spectra of the one-atom, two-atom and three-atom laser excitation probabilities $P_{1} ,\; P_{2}$, and $P_{3} $ are shown in the dimensionless detuning scale $\delta \: t_{0} $ for various average interaction pulse areas $\theta _{0} $. Laser excitation pulse area  is $\Theta =\pi $ [(a)-(c)] and $\Theta =\pi /\sqrt{3} $ (d). The F$\ddot{\mathrm{o}}$rster resonance detuning $\Delta $ is zero.}
\end{figure}

From Fig.~7 we see that the probability to excite three atoms, $P_{3} $, rapidly drops as $\theta _{0} $ increases, but probability to excite two atoms, $P_{2} $, drops much slower. This means that for the same atom density the DB does not benefit much from increasing \textit{N}. Blockade becomes stronger when we place more atoms into the same interaction volume, that is, when we increase the atom density in the mesoscopic ensemble. Another theoretical study [32] has also revealed that at $N=3$ dipole blockade cannot be complete for certain spatial configurations.

\section{CONCLUSION}

We performed the numerical Monte Carlo simulations of the Stark-tuned F$\ddot{\mathrm{o}}$rster resonance and dipole blockade between two to five cold Rydberg atoms. Various spatial configurations have been analyzed in the context of quantum-information processing with neutral atoms in optical dipole traps.

It has been shown that quantum phase gates can be implemented using coherent Rabi-like population oscillations at exact F$\ddot{\mathrm{o}}$rster resonance. However, this can be done only with well localized atoms at the temperatures below 50 $\mu {\rm K}$. Higher temperatures lead to larger atom position uncertainties and to decoherence of the population oscillations. Increasing the number of the interacting Rydberg atoms also leads to decoherence and does not help in implementing the quantum phase gates.

Dipole blockade at the Stark-tuned F$\ddot{\mathrm{o}}$rster resonance between few Rydberg atoms is more robust against the position uncertainties. The dependence of the blockade fidelity on the dipole-dipole interaction strength demonstrates a threshold-like behavior. The threshold value substantially increases if the F$\ddot{\mathrm{o}}$rster resonance detuning is not zero, but blockade is still achievable at the larger interaction strengths. In contrast, finite line width of the exciting laser radiation decreases the blockade efficiency, which cannot be improved by a stronger interaction. Increasing the number of Rydberg atoms does not increase blockade fidelity, since for the same atom density the two-atom excitation probability is almost the same as for just two atoms in the interaction volume. Finally, we have shown that dipole blockade can be observed even with two disordered atoms in a single excitation volume, if the interaction strength is large enough. 

We conclude that both quantum phase gates and dipole blockade have their advantages and drawbacks and they require different experimental conditions to be observed. In particular, quantum phase gates can be performed at a weak dipole-dipole interaction, but the atoms should be localized well to preserve coherence. Dipole blockade is insensitive to position uncertainties, but suffers from the finite laser linewidth and requires much stronger interaction, especially for quasi-F$\ddot{\mathrm{o}}$rster resonances. The results obtained can be useful for the analysis of the experimental conditions appropriate to quantum-information processing with Rydberg atoms.

\begin{acknowledgments}
This work was supported by the RFBR (Grant Nos. 09-02-90427, 09-02-92428, 10-02-00133, 10-02-92624), by the Russian Academy of Sciences, by the Presidential Grant No. MK-6386.2010.2, and by the Dynasty Foundation.
\end{acknowledgments}

\end{document}